\documentclass[useAMS,usenatbib,onecolumn]{mn2e}
\usepackage{graphicx,amsmath,epsfig,amssymb}

\title[Formation of Cyanoformaldehyde]
{Formation of Cyanoformaldehyde in the interstellar space}
\author[Ankan Das, Liton Majumdar, Sandip K. Chakrabarti, Rajdeep Saha \& Sonali Chakrabarti]
{A. Das$^{1}$, L. Majumdar$^{1}$, S. K. Chakrabarti$^{1,2}$, R. Saha$^{1}$, S. Chakrabarti$^{1,3}$ \\
$^{1}$Indian Centre for Space Physics, Chalantika 43, Garia Station Rd., Kolkata 700084, India\\
$^{2}$S. N. Bose National Centre for Basic Sciences, Salt Lake, Kolkata 700098, India\\
$^{3}$Maharaja Manindra Chandra College, 20 Ramakanta Bose lane, Kolkata 700003, India\\
}
\begin{document}

\date{}

\maketitle

\begin{abstract}
Cyanoformaldehyde (HCOCN) molecule has recently been suspected towards the Sagittarius B2(N)
by the Green Bank telescope, though a confirmation of this observation has not yet been made.
In and around a star forming region, this molecule could be formed by 
the exothermic reaction between two abundant interstellar species, H$_2$CO and CN.
Till date, the reaction rate coefficient for the formation of this molecule is unknown.
Educated guesses were used to explain the abundance of this molecule by 
chemical modeling. In this paper, we carried out quantum chemical calculations to find out 
empirical rate coefficients for the formation of HCOCN and different chemical properties 
during the formation of HCOCN molecules. 
Though HCOCN is stable against unimolecular decomposition, this gas phase molecule could be destroyed
by many other means, like: ion-molecular reactions or by the effect of cosmic rays. 
Ion-molecular reaction rates are computed by using the capture theories.
We have also included the obtained rate coefficients into our large gas-grain chemical network to
study the chemical evolution of these species in various interstellar conditions.
Formation of one of the isotopologue(DCOCN) of HCOCN is also studied. 
Our study predicts the possibility of finding HCOCN and DCOCN in the ice phase 
with a reasonably high abundance. 
In order to detect HCOCN or DCOCN in various interstellar environments, it is necessary to  
know the spectroscopic properties of these molecules. To this effect, we carried out quantum 
chemical calculations to find out different spectral parameters of HCOCN for the transition in
electronic, infrared and rotational modes. We clearly show how the isotopic substitution (DCOCN) 
plays a part in the vibrational progressions of HCOCN.

\end{abstract}


\section{Introduction}
Several complex molecules are found not only in the interstellar clouds 
but also in comets, planetary nebulae, stellar atmosphere, circumstellar envelopes etc. 
According to the Cologne Database for Molecular Spectroscopy (CDMS) catalog (Muller et al., 
(2001) \& Muller et al., (2005)) approximately 170 molecules have been detected in the 
interstellar medium or circumstellar shells. 
These molecules play an active role in the energy balance of the clouds. Despite these
overwhelmingly significant observational evidences, till date, chemical 
composition of the gas is not well resolved. Discovery of more than $20$ 
interstellar molecules in the ice phase confirmed wide spread presence of the 
interstellar dust in and around ISM. Theoretical models (Chakrabarti et al.,
2006ab; Das et al., 2008ab; Das, Acharyya\& Chakrabarti  
2010; Cuppen \& Herbst 2007; Cuppen et al., 2009; Das \& Chakrabarti 2011 etc.) indicated
that these dusts play a major role for deciding the chemical composition in
any molecular cloud. From the laboratory experiments and observational results, it is found that 
90$\%$ of the grain mantle is covered with H$_2$O, CH$_3$OH and  CO$_2$.    
The chemical composition of the gas phase of any molecular cloud is heavily dictated by 
its environment. Both the gas and the grain chemistry are influenced 
by radiation coming from nearby young stars. Several studies related to the
coupled hydro-chemical model are also present in the literature 
(e.g., Aikawa et al., 2005, Das et al., 2008a, Das et al., 2013). 
These studies explore the possibility of formation of several complex 
molecules in different regions of the molecular cloud. 

Recently, Majumdar et al., (2013) performed a quantum chemical calculation to obtain
the spectral signatures (infrared and electronic absorption spectra) of the precursors 
of some bio-molecules such as adenine, alanine and glycine. It was found that the 
spectral signature of the gas phase significantly differs from that in the ice 
phase. These type of theoretical results provide one with ideal tools for 
serving as the benchmark for the observations. Several unidentified bands 
(Unidentified Infrared emission, Allamandola, Hudgins \& Sandford, 1999) 
were identified which could be due to the hydrocarbon families. 
So, observations of some species, which are 
closely related to the hydrocarbon chains could be important clues for 
modeling purpose (Gerin et al., 1989). Formaldehyde (H$_2$CO) and HCOCN abundances could 
be strongly inter-related since they might have followed the same chemical history. 
In case of HCOCN, there is a substitution of an H atom by a CN radical. In this paper, 
we discuss different properties of HCOCN and one of its isotopologues, namely,  DCOCN.
The plan of this paper is the following: In Section 2, the models and the 
computational details are presented. Implications of the results are discussed in 
Section 3. Finally, in Section 4, we draw our conclusions. In Appendix A, we 
tabulate some of the spectral information in the format of JPL catalog.

\section{Computational details}

\begin{table*}
\centering{
\caption{Initial abundances used relative to the total hydrogen nuclei.}
\begin{tabular}{|c|c|}
Species&Abundance\\
\hline

H$_2$ &    $5.00 \times 10^{-01}$\\
He    &    $1.00 \times 10^{-01}$\\
N     &    $2.14 \times 10^{-05}$\\
O     &    $1.76 \times 10^{-04}$\\
H$_3$$^+$&    $1.00 \times 10^{-11}$\\
C$^+$ &    $7.30 \times 10^{-05}$\\
S$^+$ &    $8.00 \times 10^{-08}$\\
Si$^+$&    $8.00 \times 10^{-09}$\\
Fe$^+$&    $3.00 \times 10^{-09}$\\
Na$^+$&    $2.00 \times 10^{-09}$\\
Mg$^+$&    $7.00 \times 10^{-09}$\\
P$^+$ &    $3.00 \times 10^{-09}$\\
Cl$^+$&    $4.00 \times 10^{-09}$\\
e$^-$ &    $7.31 \times 10^{-05}$\\
\hline
\end{tabular}}
\end{table*}
\subsection{Gas phase reaction rate coefficients}

Density functional theory (DFT) is an efficient tool to explore the chemical parameters of a species. 
We use the DFT formalism explicitly to find out different chemical parameters  
for the synthesis of interstellar cyanoformaldehyde. Cyanoformaldehyde has an asymmetric 
top configuration with dipole moments along the two principal axes. It is composed of two 
functional groups (aldehyde and cyano). The computations were performed using Becke three-parameter 
Exchange and Lee, Yang and Parr correlation (B3LYP) functional (Becke, 1993) with the 6-311++G** basis 
set available in the Gaussian 09W package. Total energies, zero point vibrational energies, electronic 
energies and activation barrier energies, of all the species, which are formed during the synthesis of 
HCOCN are calculated. 

Rate coefficients for a chemical reaction could be calculated by using the conventional transition state 
theory. According to this theory, the rate coefficient has the following form:
\begin{equation}
k(T)= (K_B T/h C^0)exp(-\Delta G/R T) \ \ s^{-1},
\end{equation}
where $K_B$ is the Boltzmann constant, $h$ is the Plank's constant, $T$ is the temperature, C$^0$ is the
concentration (set to 1, following Jalbout \& Shipar, 2008), $R$ is the  ideal gas constant, and $\Delta G$ 
is the free energy of activation. It is clear from the above equation that
the rate constant is strongly dependent of temperature. 

According to Remijan et al. (2008), gas phase Cyanoformaldehyde could be formed by  
the reaction of the cyanide (CN) radical and neutral formaldehyde (H$_2$CO): 
\begin{equation*}
\mathrm{H_2CO + CN \rightarrow HCOCN + H}.
\eqno{A1}
\end{equation*}
This type of reaction is often barrier less and exothermic in nature (Balucani et al., 2002). 
H$_2$CO and CN are known to have wide-spread spatial distributions and relatively high abundances 
in the interstellar clouds (Zuckerman \& Palmer, 1974; Churchwell, 1980).         
Earlier, it was believed that HCOCN is an unstable species that would quickly undergo unimolecular 
decomposition into products of HCN and CO (e.g., Judge et al. 1986; Clouthier \& Moule 1987). 
Lewis-Bevan et al., (1992) showed that the decomposition energy for the
HCOCN is much higher (68.74 Kcal mol$^{-1} \sim 16,700K$). So once produced, 
HCOCN could survive in the interstellar environment, when shielded from the
general stellar and galactic radiation fields, i.e., in the dense cloud.

Though HCOCN is stable against unimolecular decomposition, this gas phase molecule could be destroyed
by many other interstellar chemical processes such as; ion-molecular reactions or by the effect of cosmic rays.
Since gas-phase chemistry of cold regions could be dominated by the exothermic ion-molecule reactions, 
these pathways could be served as a dominant destruction pathways for any gas phase neutral molecules. 
We have included these pathways into our gas phase chemical network for the destruction of gas phase
HCOCN.  
First of all, from Woodall et al., (2007), we have selected the major ions
in their gas phase chemical network(for this selection, steady
state abundances of the ions are shortlisted and if the abundance of any ion is found to be 
greater than $10^{-10}$ with respect to molecular hydrogen then it is selected). 
From Woodall et al., (2007), we found that abundances of 
$\mathrm{C^+,H_3O^+,HCO^+,H_3^+,Na^+,Mg^+,Fe^+,O_2^+,HN_2^+,NO^+,HNO^+ \ and \  H^+}$ were beyond 10$^{-10}$
with respect to the molecular hydrogen. Now, gas phase HCOCN 
molecules are mainly producing by the reaction A1, where one H atom of the H$_2$CO molecule is 
substituted by a CN radical. Despite of this
substitution, here we have assumed that HCOCN is interacting with the ions as the same way as H$_2$CO does. 
In between the shortlisted ions, we only have considered those ion-molecular 
destruction reaction for HCOCN molecule(from Woodall et al., 2007 network), for which the 
ion-molecular destruction pathways are available for the H$_2$CO molecule. Following are the 
list of ion-molecular reactions considered here for the destruction of gas phase HCOCN;
\begin{equation*}
H_3^+ + HCOCN \rightarrow H_2COCN^+ + H_2
\eqno{A2}
\end{equation*}
\begin{equation*}
C^+ + HCOCN \rightarrow HCO^+ + C_2N
\eqno{A3}
\end{equation*}
\begin{equation*}
C^+ + HCOCN \rightarrow CO + C_2NH^+
\eqno{A4}
\end{equation*}
\begin{equation*}
H_3O^+ + HCOCN \rightarrow H_2COCN^+ + H_2O
\eqno{A5}
\end{equation*}
\begin{equation*}
HCO^+ + HCOCN \rightarrow H_2COCN^+ + CO
\eqno{A6}
\end{equation*}
\begin{equation*}
HN_2^+ + HCOCN \rightarrow H_2COCN^+ + N_2
\eqno{A7}
\end{equation*}
\begin{equation*}
O^+ + HCOCN \rightarrow HCO^+ + OCN
\eqno{A8}
\end{equation*}
\begin{equation*}
H^+ + HCOCN \rightarrow CO^+ + HCN + H
\eqno{A9}
\end{equation*}
\begin{equation*}
H^+ + HCOCN \rightarrow HCO^+ + HCN
\eqno{A10}
\end{equation*}
According to Herbst (1996), reaction rate coefficients for the reactions that do not possess 
a potential energy barrier at short-range, long-range capture theories could be applied.
Such theories assume that all hard collisions lead to reaction and this collisions 
occur for all partial waves up to a maximum impact parameter or relative angular 
momentum quantum number.
According to them, the centrifugal barrier produces a 
long-range maximum in the effective potential energy function(V$_{eff}$):
$$
                                   V_{eff}(r,b)= v(r) +T_{AB} b^2/r^2,
$$
where, r is the separation between the reactants and $T_{AB}b^2 /r^2$ is the angular kinetic energy. If
the reactants could able to overcome the centrifugal barrier, 
they would spiral in towards each other in the absence of
short-range repulsive forces. The long-range potential(in cgs-esu units) in this situation could be 
written as follows:
$$
                                   V (r) = −e^2 \alpha_d /2 r^4 ,
$$
where $\alpha_d$ is the polarizability of the neutral reactant. This potential leads to the 
Langevin rate coefficient
\begin{equation}
                 k = v \pi b_{max}^2 =  2\pi e  \sqrt{\alpha_d / \mu}
\end{equation}
where b$_{max}$ is the maximum impact parameter, $\mu$ is the reduced mass of the reactants. 
This theory predicts a temperature independent rate coefficient with magnitude $\sim 10^{-9} cm^3 s^{-1}$.
Several experiments confirm the validity of this theory for the majority of ion-molecule reactions 
which appear rarely to possess potential energy surfaces with short range barriers.
For the case of reaction(A2-A10), HCOCN is the neutral reactant and from our quantum chemical
calculation, we get $\alpha_d=28.62$ Bohr$^3$(1 Bohr=0.529 A$^\circ$) for HCOCN. So from Eqn. 2, we could
calculate the reaction rates for the ion-molecular reactions (A2-A10) by just plugging the 
reduced mass of the respective reactants.

Gas phase ions are mainly destroyed by the Dissociative Recombination(hereafter, DR) process. 
DR pathways for the destruction of all the ions which are producing 
during the ion-molecular reactions(A2-A10), are already into our network except H$_2$COCN$^+$. 
Here, we assume that H$_2$COCN$^+$ has the similarity with H$_3$CO$^+$(one H atom of $H_3$CO$^+$ 
could be substituted by a CN radical to form H$_2$COCN$^+$). 
Following the DR pathways of the H$_3$CO$^+$, here we have considered the following
dissociative recombination pathways with the similar rate coefficient for H$_2$COCN$^+$:
\begin{equation*}
H_2COCN^+ + e^- \rightarrow CO + H_2 + CN
\eqno{A11}
\end{equation*}
\begin{equation*}
H_2COCN^+ + e^- \rightarrow HCO + H + CN
\eqno{A12}
\end{equation*}
\begin{equation*}
H_2COCN^+ + e^- \rightarrow HCOCN + H
\eqno{A13}
\end{equation*}

Interstellar species could be destroyed due the effect of cosmic rays even inside a 
dense cloud. According to Ding, Fang \& Liu (2003), HCOCN could be photo-dissociated into  
following two channels:
\begin{equation*}
\mathrm{HCOCN + h\nu \rightarrow HCO + CN},
\eqno{A14}
\end{equation*}
\begin{equation*}
\mathrm{HCOCN + h\nu \rightarrow H  +  COCN}.
\eqno{A15}
\end{equation*}
Rate of above photo-reactions (A14 \& A15) could be adopted as follows:
\begin{equation}
k_{CR}(T)= \alpha (T/300)^{\beta} \gamma/(1-\omega),
\end{equation}
where, $\alpha$ is the cosmic-ray ionization rate, $\gamma$ is the probability per cosmic-ray ionization 
that the appropriate photo reaction takes place, and $\omega$ is the dust grain albedo in the far 
ultraviolet. Following, Woodall et al., (2007), we use the cosmic ray ionization rate $\alpha$ = 
$1.30 \times 10^{-17}$ ionization per sec, $\beta = 0$, $\omega$=0.6. $\gamma$ is highly sensitive to the
reactants and thus to start with an educated estimation, unless otherwise stated, we use $\gamma=1$ for the
reactions A14 \& A15. As we are considering the gas-grain interaction, some gas phase HCOCN could 
be accreted on the grain surfaces and thus could be depleted from the gas phase. This process is 
also a depletion mechanism for HCOCN as far as the gas chemistry is concerned.

According to Nguyen and Nguyen, (1999), HCOCN could also be destroyed by the following 
reactions:
\begin{equation*}
\mathrm{HCOCN + H_2  \rightarrow H_2CO + HCN},
\eqno{A16}
\end{equation*}
\begin{equation*}
\mathrm{HCOCN + CH_4 \rightarrow HCOCH_3 + HCN},
\eqno{A17}
\end{equation*}
\begin{equation*}
\mathrm{HCOCN+ NH_3  \rightarrow HCONH_2 + HCN},
\eqno{A18}
\end{equation*}
\begin{equation*}
\mathrm{HCOCN + H2O  \rightarrow HCOOH + HCN}.
\eqno{A19}
\end{equation*}
The heat of formation for the above four reactions are
$57.2$ KJ/mol, $57.1$ KJ/mol, $55.5$ KJ/mol and $60.0$ KJ/mol respectively (Nguyen and Nguyen, 1999). 
Now, in cold interstellar clouds, averaged translational temperatures of the reactants are 
about $10$K, which can rise up to $4000$K in the outer photosphere of carbon stars 
(Kaiser et al., 1999). Since $4000$K is roughly equivalent to $40$KJ/mol, 
all the four reactions (A16-A19) do not influence the destruction
of HCOCN in the dense cloud. 

In our quantum chemical modeling, B3LYP method and 6-311++G** basis set 
are used for the calculation of the transition state for the above reactions. 
The results of the transition state calculations show that  
the reaction has a negative activation energy both in the 
gas (E$_0 = -0.825499$ eV) as well as in 
the grain (water ice) phase (E$_0= -0.609866$ eV). Any elementary reactions exhibiting 
these negative activation energies are typically barrier-less in nature and 
the reactions relie on the capture of the molecules in a potential well. 
Reaction rate coefficient for such a reaction A1 could be calculated by using Eqn. 1.
For the gas phase reaction A1, Quantum chemical calculation 
during the transition state provides, $\Delta G$. (Free energy of activation with the thermal 
correction)$=1.60381843 $ Kcal/mole. 
Here, we restrict our simulations in the temperature range of $10-20$K, which is feasible 
for the dense cloud condition. It is clear from the rate expression (Eqn. 1) that it 
has an exponential dependency on the temperature. During this temperature range ($10-20$K),
the gas phase Rate coefficients for reaction A1 varies from $1.838 \times 10^{-24} s^{-1}$ to
$1.237 \times 10^{-06} s^{-1}$.

\subsection{Surface reaction rate coefficients}
We have assumed that the gas phase species are physisorbed onto the dust 
grains ($\sim 1000$ A$^\circ$) 
having a grain number density of $1.33 \times 10^{-12}n_H$, 
where $n_H$ is the concentration of H nuclei in all forms.   
Binding energies of the surface species are the keys for the chemical 
enrichment of the interstellar grain mantles. Grain surface
provides the space for the interstellar gas phase species 
to land on and to react with other surface species. Chemical species 
can return back to the gas phase after chemical reactions or in the 
original accreted form. Reactions among
the surface species are highly dependent on their mobility
which in turn, depends on the thermal hopping
time scale or on the tunneling time scale whichever is shorter.
For the lighter species, the tunneling time scale is 
much shorter than the hopping time scale. 
When two surface species meet, they can react. If some activation energy 
is required for that reaction to happen, the reaction occur with a quantum mechanical
tunneling probability. Following Hasegawa, Herbst \& Leung (1992), reaction
rate (R$_{ab}$) between the surface species $a$ and $b$ can be expressed as;
\begin{equation}
R_{ab}=k_{ab}(R_{d,a}+R_{d,b})N_a N_b n_d,
\end{equation}
where, N$_a$ and N$_b$ are the the number of species `a' and `b' 
on an average grain respectively, k$_{ab}$ is the probability 
for the reaction to happen upon an encounter and n$_d$ is the 
dust-grain number density, R$_{d,a}$ and R$_{d,b}$ are the
diffusion rate respectively for the species `a' and `b'.
Diffusion rate (R$_{d,a}$ =$\frac{1}{N_St_{d,a}})$ depends upon the time 
needed to traverse entire grain by the reactive species, which in turn depends on
\begin{equation}
t_{d,a}=\nu_0^{-1} exp(E_b/kT_g) sec,
\end{equation}
where, $\nu_0$ is the characteristic vibration frequency of the adsorbed
species, E$_b$ is the potential energy barrier between adjacent 
surface potential energy wells, T$_g$ is the grain temperature and N$_S$ is the number 
of surface sites on a grain. For the computation of the characteristic vibration frequency ($\nu_0$),
we have utilized the following harmonic oscillator relation:
\begin{equation}
\nu_0=\sqrt{(2 n_sE_d/\pi^2 m)},
\end{equation}
where, $n_s$ is the surface density of sites, m is the mass of the adsorbed particle and $E_d$ is the
adsorption energy of the species.
Here we use N$_S$=10$^6$ and $n_s=2 \times 10^{14}$.
	
Ice phase abundance of the surface species could be decreased by (a) thermal evaporation,
(b)cosmic ray induced photo-dissociation and (c) cosmic ray induced evaporation process. 
Rate of thermal evaporation of the surface species `i' could be calculated by the following relation;
\begin{equation}
k_{evap}(i)=\nu_0 exp(-E_d/kT_g) \ \ sec^{-1},
\end{equation}
where, $E_d$ is the adsorption energy of the i$^{th}$ species. 

Following the Photo-dissociation channels of gas phase HCOCN, we have assumed that ice phase
HCOCN \& DCOCN could also be dissociated into the similar ways. So following, 
reaction A14 \& A15, photo-dissociation channel of the ice phase DCOCN molecule could be as follows:
\begin{equation*}
\mathrm{DCOCN + h\nu \rightarrow DCO + CN},
\eqno{A20}
\end{equation*}
\begin{equation*}
\mathrm{DCOCN + h\nu \rightarrow D  +  COCN}.
\eqno{A21}
\end{equation*}
Moreover, we have assumed that Photo-dissociation rate of the surface species are same as in 
the gas phase (Eqn. 3).

Rate of cosmic ray induced evaporation are 
calculated by using the expression developed by Hasegawa \& Herbst (1993). 
Hasegawa \& Herbst (1993) did their simulation for gas and grain 
temperature 10K and for hydrogen number density = $2 \times 10^4$ cm$^{-3}$ cloud. 
Following Leger et al., (1985), they assumed that relativistic Fe nuclei with 
energies 20-70MeV could deposit 0.4 MeV energy on an average dust particle of 
radius 0.1$\mu m$. Now grain could be cooled down due to the thermal evaporation and radiation process. 
For the easy inclusion of cosmic ray induced photo-evaporation into their model, 
they developed the following relation:
\begin{equation}
k_{crd}\sim f(70,K) k_{evap}(i,70 K),
\end{equation}
where, $k_{evap}(i, 70K)$ is the thermal evaporation rate of the surface species `i' at temperature 70K, 
f(70 K) is the fraction of the time spent by grains in the vicinity of
70K. Following Leger et al., (1985), they defined $f(70 K)=3.16 \times 10^{-19}$. 

In our surface network, we have considered that the following reactions along with reaction number A1
for the production of HCOCN and DCOCN in the grain phase;
\begin{equation*}
\mathrm{HDCO + CN \rightarrow DCOCN + H},
\eqno{A22}
\end{equation*}
\begin{equation*}
\mathrm{HDCO + CN \rightarrow HCOCN + D},
\eqno{A23}
\end{equation*}
\begin{equation*}
\mathrm{D_2CO + CN \rightarrow DCOCN +D}.
\eqno{A24}
\end{equation*}
Reaction rate coefficients for the surface reactions (A1, A22, A23 \& A24) are calculated by using
Eqn. 3. Required adsorption energies for this calculations are mainly taken from 
Allen \& Robinson (1977) and Hasegawa \& Herbst (1993). Following Hassegawa, Herbst \& Leung (1992), 
here also, we have assumed that E$_b$=0.3E$_d$ except the case of atomic hydrogen. 
As like Hasegawa, Herbst \& Leung (1992), here also 
we use, E$_b$=100K for atomic hydrogen. Binding energy of the deuterated species are assumed to be
same as its hydrogenated counter part.

\subsection{Spectral parameters}

Quantum chemical calculations could be very accurate for the identification of several 
species in the interstellar medium. Huang \& Lee (2008) proved that quantum chemical calculations 
might provide rotational constants often up to an accuracy of $20$ MHz 
(especially for the B and C type constants) and also vibrational
frequencies accurate to 5 cm$^{-1}$ or better (Huang \& Lee 2008, 2009, 2011; Huang et al. 2011;
Inostroza et al. 2011; Fortenberry et al. 2011a, 2011b, 2012a, 2012b, 2012c). Following this type of
earlier works, we are motivated to present the spectroscopic constants as well as the
fundamental vibrational frequencies to assist in confirming detections of HCOCN and DCOCN in 
and around an ISM.

In order to study the spectroscopy of cyanoformaldehyde, we first optimize 
the geometry of the cyanoformaldehyde molecule at density functional theory based 
B3LYP method using 6-311++G** basis set. Vibrational frequencies of cyanoformaldehyde 
and one of its isotopomer is computed by determining the second derivative of the 
energy with respect to the Cartesian nuclear coordinates and then transforming 
into mass-weighted coordinates. This transformation is valid only 
at a stationary point. Here, we are mainly considering the gas phase and the 
ice phase cyanoformaldehyde and one of its isotopomers. Properties of molecules 
and transition states can differ considerably between 
the gas phase and the solution phase. For example, the electrostatic effects are often much less important
for species placed in a solvent with high dielectric constant than when they are 
in the gas phase (Foresman \& Frisch, 1996). In our study, we consider simple as well as mixed
ice. Simple ice consists of only water but the observational 
evidences suggest that $\sim 90$\% interstellar grain mantle 
is covered with H$_2$O, CH$_3$OH and CO$_2$ (Keane et al., 2001). This is why we also 
find the vibrational frequencies of these species by considering the actual percentage 
of the mixed solvent. The influence of the solvent in the vibrational 
spectroscopy of these species is done using the Polarizable 
Continuum Model (PCM) with the integral equation formalism variant (IEFPCM) 
as a default Self-consistent Reaction Field (SCRF) method. 
Among the different models, we have chosen IEFPCM model as
a convenient one, since the second energy derivative is 
available for this model and also its analytic form is available. We also 
find the electronic absorption spectrum of HCOCN(for gas phase, ice phase and mixed ice phase) 
using the time dependent density functional theory (TDDFT study by IEFPCM model).

Rotational motion is commonly described by starting with the rigid rotor model. Calculations 
at the rotational level requires very higher level of basis sets for the better estimation
of the structure and optimization. Here we use  MP2/aug-cc-pVTZ level for performing our 
calculations. Some DFT methods with larger basis set could also be used as well.
Corrections for the interaction between rotational motion and
vibrational motion along with the corrections for vibrational averaging and
an-harmonic corrections to the vibrational motion are also considered in our calculations. 
In brief, we report rotational constants for HCOCN, which are corrected for 
each vibrational state as well as vibrationally averaged structures. 
These rotational constants are needed to predict the spectrum of cyanoformaldehyde molecule and this
can be done using the `SPCAT' program (Pickett 1991). The SPCAT program requires two 
main files namely, `file.var' and `file.int'. The `.var' file is generated from quantum chemical 
simulations by using Gaussian 09W software. This file contains
information about rotational constants, quadrupole coupling constants and distortion constants.
The `.int' file is the intensity file and it is prepared according to the prescribed format
of `SPCAT program'. This file contains the maximum and minimum rotational states,
partition function, rotational temperature and the dipole moment of the molecule.
 
\subsection{Chemical Modeling}
\begin{figure}
\vskip 1cm
\centering
\includegraphics[height=11cm,width=13cm]{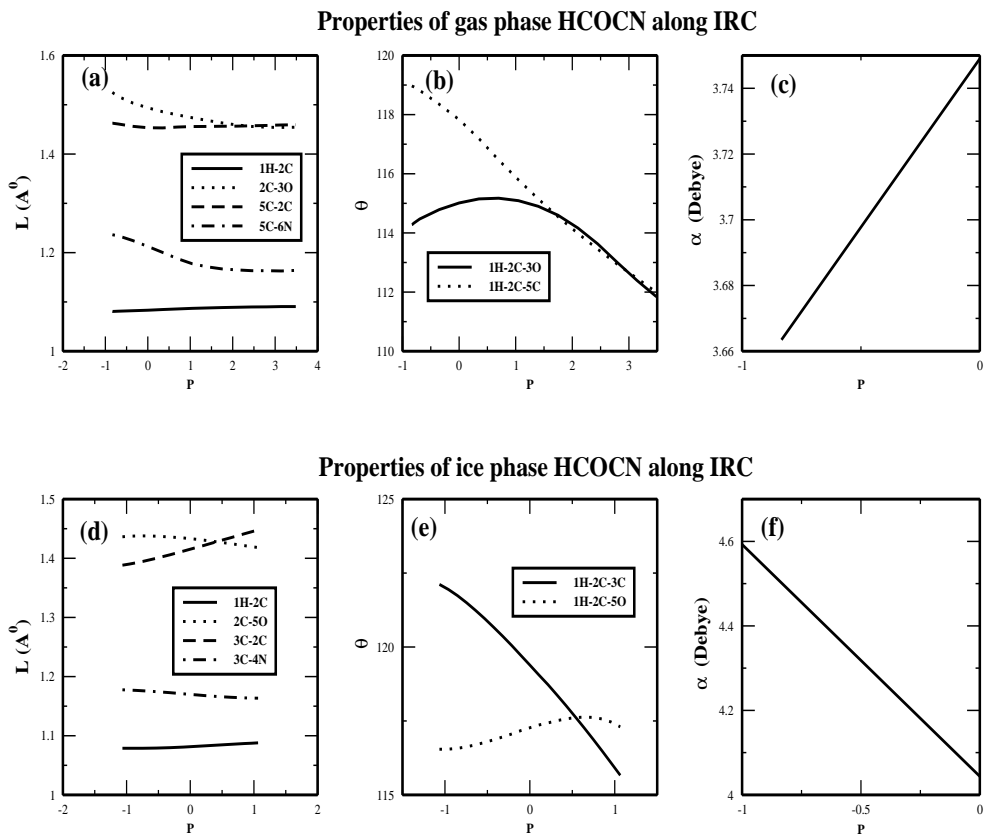}
\caption{Variation of different Molecular properties for the formation of HCOCN in gas phase (a-c)
and ice phase (d-f) with respect to the intrinsic reaction coordinate.}
\end{figure}

Our chemical model consists of a gas phase chemical network as well as the surface chemical network.
Our gas phase chemical network consists of 5229 reactions among 591 species. 
This large gas phase network includes the network of Woodall (2007).
In addition, we have introduced the formation and destruction pathways of 
cyanoformaldehyde, some important deuterated reactions 
following Robert \& Millar, (2000) \&  Albertsson et al., (2011),
and some reactions which lead to the formation of bio-molecules by following
Chakrabarti \& Chakrabarti (2000ab) \& Majumdar et al. (2012). 
We assume that the gas and the grains are coupled through the accretion and thermal evaporation processes. 

For the production of HCOCN on the grain surface, reactions A1 \& A23 are considered. 
Along with these surface reactions, a large surface chemical network
are used. Our surface chemical network consists of 285 surface reactions among 154 surface species.
For this surface chemical network, we mainly have followed Hasegawa, Herbst \& Leung (1992);
Das et al., (2008); Das, Acharyya \& Chakrabarti (2010); Das \& Chakrabarti., (2011); Cazaux et al., (2010);
Cuppen \& Herbst (2007) \& Stantcheava et al., (2002).
Binding energies of all the surface species are mainly taken from Allen \& Robinson (1977)
and Hasegawa \& Herbst (1993).
Binding energies of the deuterated species are assumed to be the same as its hydrogenated counter part.
Here, we use parameters which are suitable to mimic
a cold and dense interstellar cloud, namely $T = 10-20$K, n$_H$ = $2 \times 10^4$ cm$^{−3}$,
A$_V=10$ magnitude. Following Lee et al., (1996), in Table 1, initial fractional abundances
relative to the total hydrogen nuclei is shown.
This type of initial abundances are often adopted for the cold \& dark cloud.
Initial abundance of
`D' in the gas phase is varied from a concentration n$_D$ = 0.001 cm$^{-3}$ to
10 cm$^{-3}$. This implies an initial deuterium fractionation  of
(atomic D/H ratio, hereafter R$_D$) $0.001-10$. Unless otherwise stated, following Caselli et al.,
(2002), we use R$_D=0.3$.

\begin{figure}
\vskip 1cm
\centering
\includegraphics[height=8cm,width=8cm]{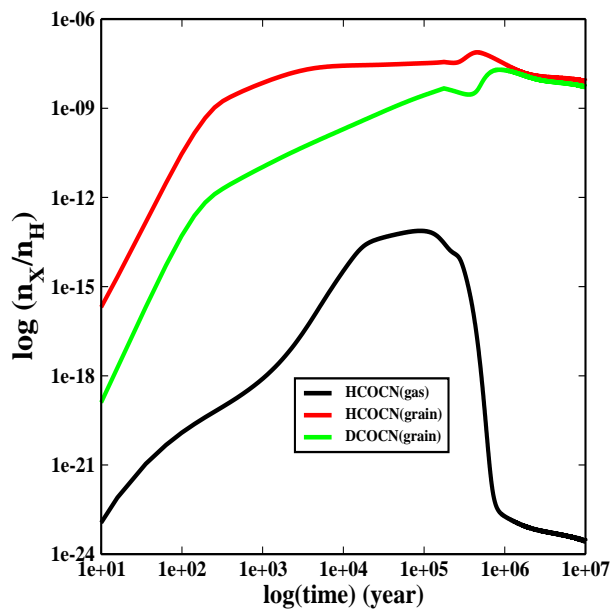}
\caption{Time evolution of HCOCN and one of its isotopomer DCOCN in gas as well as in grains.}
\label{fig-1}
\end{figure}
\begin{table*}
\scriptsize
\centering
\vbox{
\caption{Energies of the reactants, products and activated complex for the reaction H$_2$CO + CN$\rightarrow$ HCOCN + H in  the gas and ice phase.}
\begin{tabular}{|c|c|c|c|c|}
\hline
{\bf Species}&{\bf Gas phase SCF energy }&{Gas phase ZPE}&{\bf Ice phase SCF energy }&{\bf Ice phase ZPE}\\
&(Hartree)&(Hartree)&(Hartree)&(Hartree)\\
\hline
{\bf H$_2$CO}&-114.500430&0.026623&-114.507313&0.026722\\
{\bf CN}&-92.707565&0.004675&-92.709497&0.004680\\
{\bf TS}&-207.242371&0.035337&-207.239089&0.031268\\
{\bf HCOCN}&-206.728293&0.026326&-206.737219&0.026364\\
{\bf H}&-0.502257&0.000000&-0.502283&0.000000\\
\hline  
\end{tabular}}
\end{table*}
\begin{figure}
\vskip 1cm
\centering
\includegraphics[height=8cm,width=8cm]{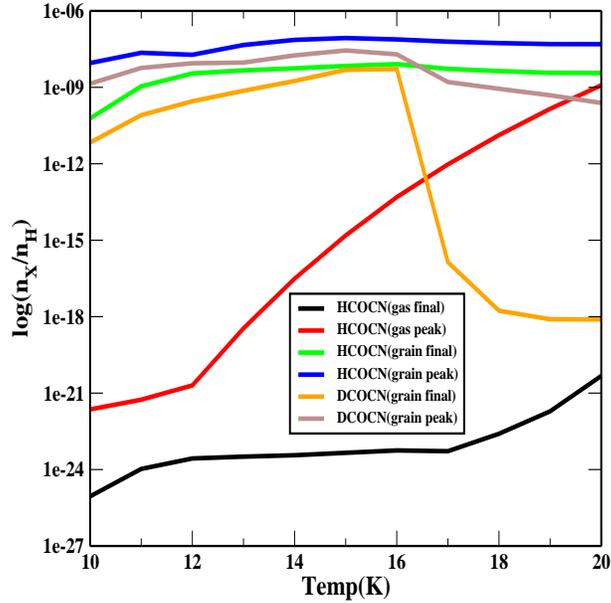}
\caption{Variation of abundances with temperature.}
\label{fig-2}
\end{figure}
\begin{figure}
\vskip 1cm
\centering
\includegraphics[height=8cm,width=8cm]{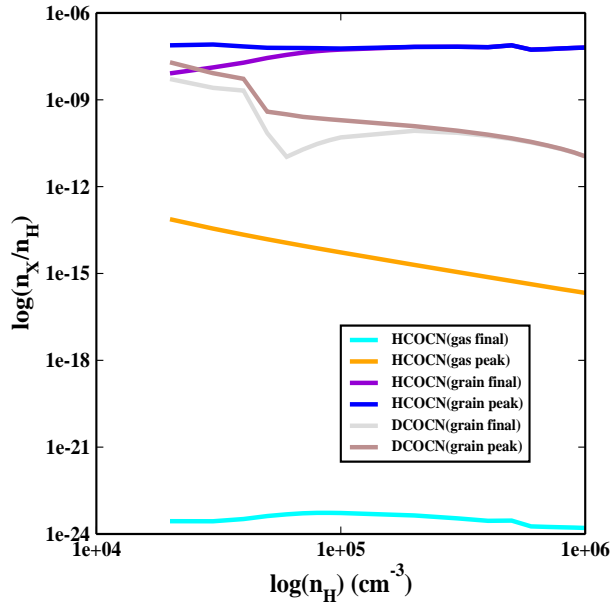}
\caption{Variation of abundances with Hydrogen number density at $16$K.}
\label{fig-2}
\end{figure}
\begin{figure}
\vskip 1cm
\centering
\includegraphics[height=8cm,width=8cm]{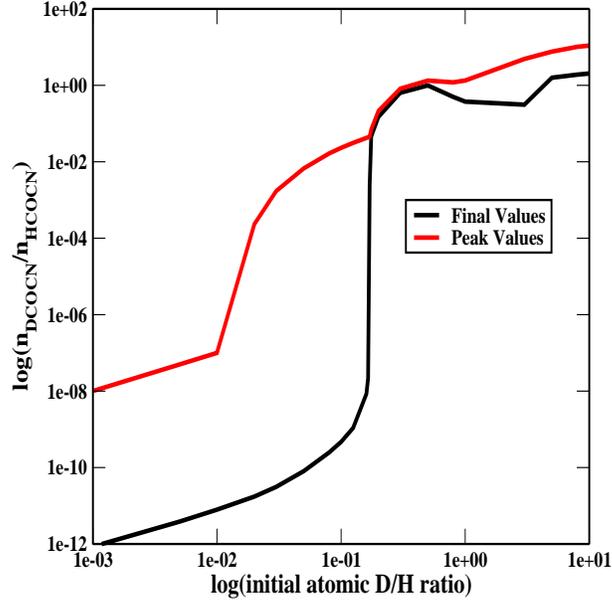}
\caption{Deuterium fractionation of HCOCN. as a function of the initial D/H ratio.}
\label{fig-2}
\end{figure}

\begin{table}
\centering
\addtolength{\tabcolsep}{-4pt}
\caption{Rate coefficients}
\begin{tabular}{|c|c|c|c|}
{\bf Reaction}&{\bf Temperature (in K)}&{\bf Gas phase reaction rates \& type }&{\bf Surface reaction rates}\\
\hline
$\mathrm{H_2CO+CN \rightarrow HCOCN +H}$(A1)&16&$4.11 \times 10^{-11}$ sec$^{-1}$(Radical-neutral)&$6.16 \times 10^{-7}$ sec$^{-1}$\\
\hline
$\mathrm{H_3^+ + HCOCN \rightarrow H_2COCN^+ + H_2}$(A2)&-&$2.85 \times 10^{-9}$ cm$^3$sec$^{-1}$(Ion-molecular)&\\
\hline
$\mathrm{C^+ + HCOCN \rightarrow HCO^+ + C_2N}$(A3)&-&$1.53 \times 10^{-9}$ cm$^3$sec$^{-1}$(Ion-molecular)&\\
\hline
$\mathrm{C^+ + HCOCN \rightarrow CO + C_2NH^+}$(A4)&-&$1.53 \times 10^{-9}$ cm$^3$sec$^{-1}$(Ion-molecular)&\\
\hline
$\mathrm{H_3O^+ + HCOCN \rightarrow H_2COCN^+ + H_2O}$(A5)&-&$1.28 \times 10^{-9}$ cm$^3$sec$^{-1}$(Ion-molecular)&\\
\hline
$\mathrm{HCO^+ + HCOCN \rightarrow H_2COCN^+ + CO}$(A6)&-&$1.1 \times 10^{-9}$ cm$^3$sec$^{-1}$(Ion-molecular)&\\
\hline
$\mathrm{HN_2^+ + HCOCN \rightarrow H_2COCN^+ + N_2}$(A7)&-&$1.1 \times 10^{-9}$ cm$^3$sec$^{-1}$(Ion-molecular)&\\
\hline
$\mathrm{O^+ + HCOCN \rightarrow HCO^+ + OCN}$(A8)&-&$1.36 \times 10^{-9}$ cm$^3$sec$^{-1}$(Ion-molecular)&\\
\hline
$\mathrm{H^+ + HCOCN \rightarrow CO^+ + HCN + H}$(A9)&-&$4.85 \times 10^{-9}$ cm$^3$sec$^{-1}$(Ion-molecular)&\\
\hline
$\mathrm{H^+ + HCOCN \rightarrow HCO^+ + HCN}$(A10)&-&$4.85 \times 10^{-9}$ cm$^3$sec$^{-1}$(Ion-molecular)&\\
\hline
$\mathrm{H_2COCN^+ + e^- \rightarrow CO + H_2 + CN}$(A11)&-&$2 \times 10^{-7}$ cm$^3$sec$^{-1}$(Dissociative Recombination)&\\
\hline
$\mathrm{H_2COCN^+ + e^- \rightarrow HCO + H + CN}$(A12)&-&$2 \times 10^{-7}$ cm$^3$sec$^{-1}$(Dissociative Recombination)&\\
\hline
$\mathrm{H_2COCN^+ + e^- \rightarrow HCOCN + H}$(A13)&-&$2 \times 10^{-7}$ cm$^3$sec$^{-1}$(Dissociative Recombination)&\\
\hline
$\mathrm{HCOCN + h\nu \rightarrow HCO + CN}$(A14)&-&$3.25 \times 10^{-17}$ sec$^{-1}$(Photo-dissociation)&$3.25 \times 10^{-17}$ sec$^{-1}$\\
\hline
$\mathrm{HCOCN + h\nu \rightarrow H + COCN}$(A15)&-&$3.25 \times 10^{-17}$ sec$^{-1}$(Photo-dissociation)&$3.25 \times 10^{-17}$ sec$^{-1}$\\
\hline
$\mathrm{DCOCN + h\nu \rightarrow DCO + CN}$(A20)&-&$3.25 \times 10^{-17}$ sec$^{-1}$(Photo-dissociation)&$3.25 \times 10^{-17}$ sec$^{-1}$\\
\hline
$\mathrm{DCOCN + h\nu \rightarrow D + COCN}$(A21)&-&$3.25 \times 10^{-17}$ sec$^{-1}$(Photo-dissociation)&$3.25 \times 10^{-17}$ sec$^{-1}$\\
\hline
$\mathrm{HDCO + CN \rightarrow HCOCN + D}$(A22)&16&-&$6.16 \times 10^{-7}$ sec$^{-1}$\\
\hline
$\mathrm{HDCO + CN \rightarrow HCOCN + D}$(A23)&16&-&$6.16 \times 10^{-7}$ sec$^{-1}$\\
\hline
$\mathrm{D_2CO + CN \rightarrow DCOCN +D}$(A24)&16&-&$6.16 \times 10^{-7}$ sec$^{-1}$\\
\hline
\end{tabular}
\end{table}

\section{Results and Discussions}

\subsection{Chemical properties}
In order to compute the energy of the transition structure of the reaction A1,
we mainly perform three sets of calculations. First, we obtain the self-consistent field (SCF) energy
by optimizing the transition structure geometry. Second, to obtain the zero point energy,
we perform the frequency calculations and finally, we perform an 
Intrinsic Reaction coordinate (IRC) calculation. 
An IRC calculation examines the reaction 
path leading down from a transition structure on a potential energy
surface and this is used as a composite variable, which spans all the degrees 
of freedom of the potential energy surface. 
Such a calculation starts at the saddle point and follow the path in both the 
directions from the transition state by optimizing the geometry of the molecular 
system at each point along the path. 
Thus an IRC calculation connects two minima on the potential energy surface 
by a path which passes through the transition state between them.
The energy of the reactants, products and activated complex with their corresponding 
zero point corrected values (in the unit of Hartree, 1 Hartree=$27.211$eV) are given in Table 2.

Fig. 1(a-f) shows the variation of different molecular parameters during the formation of HCOCN
with respect to the intrinsic reaction co-ordinate (P)
for the gas phase (H$_2$CO+CN$\rightarrow$H(1)C(2)O(3)C(5)$\equiv$N(6)+H) as well as for the
ice phase (H$_2$CO+CN$\rightarrow$H(1)C(2)O(5)C(3)$\equiv$N(4)+H) by the reaction A1.
Variation of the molecular parameters are shown in order to have the idea of the effect of solvent
during any chemical reaction.  
Numbers in the parenthesis indicate the positions of the atoms during
the reaction A1 in the gas phase as well as in the ice phase.
Reaction paths of any chemical reactions are connected by the reactants and the
products through its transition state. Fig. 1(a-c) shows the
variation of bond length (L in A$^{\circ}$), bond
angle ($\theta$ in degree) \& dipole moment ($\mu$ in Debye) along the IRC
of the gas phase formation of HCOCN. Fig. 1(d-f) represent similar
variations along IRC for the formation of HCOCN in the ice phase.
In Fig. 1a, we show how the bond lengths (1H-2C, 2C-3O, 5C-2C \& 5C-3O)
vary. It is clear from the Figure that 5C-3O bond shows strong variation along the IRC.
In case of Fig. 1d (ice phase HCOCN), we show similar variations for the
1H-2C, 2C-5O, 3C-2C \& 3C-5O bond lengths. Bond lengths in the ice phase do not show any
strong variation along IRC. In Fig. 1b, variations of bond angles (1H-2C-3O \& 1H-2C-5C) are
shown for the gas phase and in Fig. 1e, variations of bond angles (1H-2C-3C \& 1H-2C-5O) are
shown for the ice phase. In Fig. 1c  and Fig. 1f,
variations of Dipole moments are shown for the gas and ice phase respectively.
Dipole moment for the gas phase shows a strong increasing slope for the formation of
HCOCN, whereas for the ice phase formation of HCOCN, it shows a strong
decreasing slope along IRC.  
Increasing slope of the dipole moment in the gas phase could be explained due to
a substitution of the H atom by a more electronegative group (-CN) whereas a
decreasing slope of the ice phase could be explained due to the
increase of the  of 3C-2C bond and the increase of
$\angle$1H-2C=5O bond angle (Fig.1d  and Fig.1e respectively) and due to the strong interaction
with the solvent.

Calculation of gas phase rate coefficients are already discussed in Section 2.1.
Rate coefficients for the gas phase reaction A1 are calculated by using free energy of
activation (using Eqs. 1) and the rate coefficients for the ion-molecular reactions(A2-A10)
are calculated using the capture theories(Eqn. 2). In Table 3, we have summarized all the 
calculated/assumed rate coefficients for the formation/destruction of HCOCN and DCOCN in gas/ice phase.

\subsection{Chemical evolution \& deuterium enrichment}
Rate coefficients for the surface reactions are calculated by following the procedure discussed in the 
Section 2.2. 
According the discussion above, binding energies are the key for the surface reactions.
For the production of HCOCN, it is essential to have a knowledge about the binding energies of the
reactants with the grain surface.
Following Allen \& Robinson (1977), we use adsorption energies of H$_2$CO and CN
3.5 Kcal mole$^{-1}$ and 3 Kcal mole$^{-1}$ respectively.
Adsorption energies of HDCO and D$_2$CO are assumed to be similar
to the H$_2$CO molecule. In Table 3, the calculated rate coefficients are tabulated in sec$^{-1}$ for 
16K cloud. Now in order to have an idea about the abundances of the HCOCN and DCOCN around the
dense interstellar cloud, we ran several cases and all the relevant abundances are presented 
with respect to the total hydrogen.
In Fig. 2, the time evolution of HCOCN abundance in the gas phase \& ice phase along with the
DCOCN abundance in the ice phase are shown for T=16K, A$_V$=10, n$_H=2 \times 10^4$ cm$^{-3}$ and R$_D=0.3$. 
\begin{figure}
\vskip 1cm
\centering
\includegraphics[height=8cm,width=8cm]{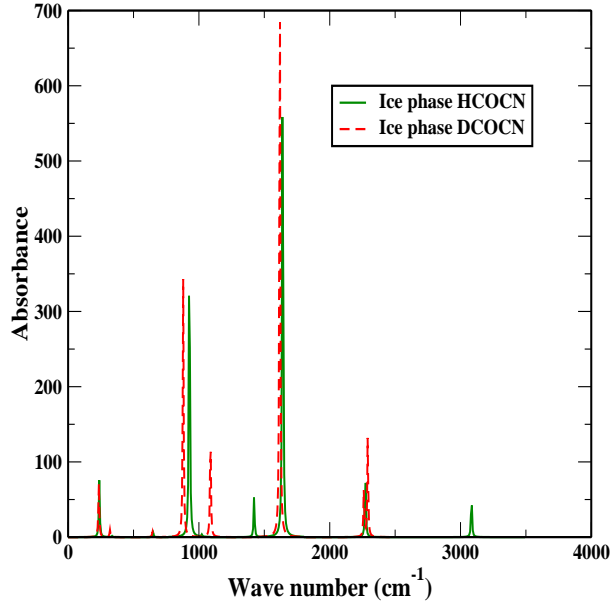}
\caption{Infrared spectrum of HCOCN and DCOCN in water ice.}
\label{fig-2}
\end{figure}

In the gas phase, production of HCOCN molecule is significant but
HDCO or D$_2$CO, which are required to form DCOCN (Eqn. A22 \& A24) are not
abundant in the gas phase. Due to this reason, we do not consider the production of
DCOCN in our gas phase chemical network. In the ice phase, HDCO or D$_2$CO could
be produced efficiently, which then react with CN radical to form DCOCN (reaction A22 and A24).
From Fig. 2, it is clear that HCOCN could be efficiently produced in the gas phase having peak abundance
$7.51 \times 10^{-14}$. It is distinctly clear from Fig. 2, that HCOCN and DCOCN both are 
significantly abundant on the grain surface having peak abundance $7.61 \times 10^{-8}$ and 
$1.98 \times 10^{-8}$ respectively.

Here, we mainly concentrate on the temperature range of $10-20$K to
mimic the dense cloud condition. Due to the exponential dependency of 
the gas phase rate coefficient (Eqn. 1) on temperature,
the rate coefficient sharply rises with temperature of the gas. This,
in turn, increases the production of gas phase HCOCN. In Fig. 3, the peak abundances of the gas phase HCOCN,
ice phase HCOCN and DCOCN as well as the final abundances (by final abundance, we want to mean 
the abundance after the life time $\sim 10^7$ years of a molecular cloud) of these species
with respect to the temperature are shown. For this case, we consider $n_H=2 \times 10^4 cm^{-3}$,
$A_V=10$ and vary the temperature from 10K to 20K. Abundance of the gas phase 
HCOCN (peak values as well as the final values) increases sharply due 
to the enhancement of the rate coefficient with temperature. Its peak abundance varies 
from $2.3 \times 10^{-22}$ to $1.21 \times 10^{-9}$ as the temperature 
is increased from 10K to 20K, whereas the final abundance of gas phase HCOCN varies from 
$8.84 \times 10^{-26} \ cm^{-3}$ to $4.72 \times 10^{-21}$ in between the specified temperature range. 
Surface production of HCOCN or DCOCN depends upon the surface population of H$_2$CO, HDCO, D$_2$CO and CN
molecules. As the temperature increases, the thermal hopping time scale (Eqn. 5) would be much shorter, which
makes the sweeping rate to be much faster. But as the thermal evaporation time scale also decreases with the
temperature, residential time of the H or D atoms on the grain surface decreases. Moreover, there are always
a competition going on between the H and the D atoms. As we have considered R$_D=0.3$ for this case, 
abundance of H atom is a bit higher compare to the D atom, that is why final abundance of HCOCN does not 
show a drastic drop as in the case of the final surface abundance of DCOCN (Fig. 3).
The peak values of the ice phase HCOCN abundance oscillates between 
$8.97 \times 10^{-9}-4.94 \times 10^{-8}$ and the peak values of the ice phase DCOCN abundance 
oscillates between $1.40 \times 10^{-9}-2.44 \times 10^{-10}$. Whereas, 
the final values of ice phase HCOCN oscillates between $6.07 \times 10^{-11}-3.67 \times 10^{-9}$ and 
the final values of ice phase DCOCN oscillates between $6.83 \times 10^{-12}-7.80 \times 10^{-19}$.

Remijan et al., (2008) made a few simple assumptions while computing the HCOCN abundances. 
According to them, the gas phase HCOCN abundance could be 4 $\times 10^{-9}$ 
with respect to the total hydrogen.
Following the work of Lewis-Bevin et al., (1992) they considered that Cyanoformaldehyde could be destroyed
by reacting with H$_2$O molecules. Assuming that the formation of Cyanoformaldehyde is dominated mainly 
by the reaction of CN with H$_2$CO with a rate of $k_1$, and that its destruction is due 
to the reaction with H$_2$O with a rate $k_2$, they determined the relative
cyanoformaldehyde abundance ratio as $X(Cyanoformaldehyde) \sim k_1  X(CN)  X(H_2CO)/ {k_2  X(H_2O)}$;
with $X(CN)\sim X(H_2CO)\sim 2 \times 10^{-8}, X(H_2O) \sim 1 \times 10^{-7}$, and $k_1\sim k_2$, 
$X(CNCHO) ~ 4 \times 10^{-9}$. In our case, we mentioned that the heat of formation for the reaction between 
Cyanoformaldehyde and water (reaction A5) is 60 KJ/mole, so this reaction is not feasible around the
dense cloud region. Moreover Remijan et al., (2008) considered that the formation rate of
Cyanoformaldehyde($k_1$) is equal to the destruction rate ($k_2$). This assumption could be 
very inaccurate.
Remijan et al.,(2008) also derived the column density of HCOCN from the data derived by Martin et al., (2004).
According to their calculations, HCOCN abundance would be in the range 
$(0.7-11) \times 10^{-9}$ for a molecular hydrogen column density of 1.6 $\times 10^{23}$ cm$^{-2}$ and
column density would be in the range of $(1-17) \times 10^{14} \ cm^{-2}$. 

Following Shalabiea et al., (1994) and Das \& Chakrabarti (2011), the column density of a species could be
calculated by the following relation;
\begin{equation}
N(A)=n_H x_i R,
\end{equation}
where, n$_H$ is the total hydrogen number density, x$_i$ is the abundance of the i$^{th}$
species and $R$ is the path length along the line of sight (=$\frac{1.6 \times 10^{21} \times A_V}{n_H}$).

In Fig. 4, we show the abundance variation of the gas phase HCOCN, ice phase HCOCN and ice phase DCOCN
at $T=16$K with respect to the variation of the number density of the cloud
in between n$_H$=2$\times 10^4$ cm$^{-3}-10^6$cm$^{-3}$. Gas phase abundance of HCOCN
shows a decreasing trend with the increase in the number density
of the cloud. Peak value of the gas phase HCOCN decreases from $7.51 \times 10^{-14}$  to
$2.12 \times 10^{-16}$ for the change of number density from $2 \times 10^4 \ cm^{-3}$ to
$10^6 \ cm^{-3}$. From Eqn. 9, these abundances result in the peak column density of 
$1.20 \times 10^9 \ cm^{-2}$ to $3.39 \times 10^6 \ cm^{-2}$, which is beyond the observational limit. 
The behaviour of the grain phase production is more interesting.
Final value of ice phase HCOCN shows an increasing trend up to $n_H \sim 10^5 \ cm^{-3}$.
Beyond that, the abundances remain roughly constant even when 
the number density of the cloud is raised by an order of magnitude. 
Beyond $10^5 \ cm^{-3}$, final abundance of ice phase HCOCN 
and peak value roughly overlaps. In between, the density range considered here, 
the peak value of the ice phase HCOCN fluctuates between $7.61 \times 10^{-8}$ to $6.38 \times 10^{-8}$,
which corresponds to the column density of $1.22 \times 10^{15} \ cm^{-2}$ to $1.02 \times 10^{15} \ cm^{-2}$.
In case of ice phase DCOCN, peak value decreases with the increase in the 
number density of the cloud, whereas initially decreasing trend was observed for for the final values of 
DCOCN. Beyond $2 \times 10^5$ year, final values and peak values follow the same decreasing trend.
Peak value of the ice phase DCOCN for $2 \times 10^4 \ cm^{-3}$ cloud was $1.98 \times 10^{-8}$ 
which reaches $1.12 \times 10^{-11}$ for the $10^6 \ cm^{-3}$ number density cloud. This implies 
a column density of $3.17 \times 10^{14}-1.79 \times 10^{11} \ cm^{-2}$.

Around the low extinction region (A$_V$$<$5), the chemical composition is heavily affected by the
interstellar photon (Das \& Chakrabarti, 2011). Das \& Chakrabarti (2011), also showed that
if we assume that the number density of the cloud does not vary with the extinction
for A$_V > 10$ then the ice composition does not vary at all. Now if we considered, say,
A$_V \sim 145$ (a value considered by Allamandola et al., 1992 for W33A) 
for the case of Fig. 4, then the computed abundances would not show any significant variations.
But while we would transform this abundances into column densities by Eqn. 8 there
would be an increase in the column densities by a factor of $14.5$. So, in between the density range 
$2 \times 10^4$ to $10^6$ cm$^{-3}$, T=16K and for A$_V$=145, a peak column density of gas 
phase HCOCN comes out to be $1.74 \times 10^{10} \ cm^{-2}$, whereas peak column densities of the 
ice phases of HCOCN and DCOCN come out to be $1.77 \times 10^{16}$ cm$^{-2}$ and 
$4.6 \times 10^{15}$ cm$^{-2}$
respectively. It is well known that if some energetic events occur, 
then most of the surface species would get lost from the surface and populate 
the gas phase. Since here we are considering that the temperature remains 
the same ($T=16$K) in the simulation time scale, most of the complex surface species once 
produced/accreted get trapped into the potential well of the grain surface. 
Only some percentage of the species could be evaporated  
by the thermal evaporation (Eqn. 5) or cosmic ray induced evaporation process (Eqn. 6). But
these are not efficient enough to populate the gas phase. Our calculated ice phase column densities are 
in line with the gas phase abundance prediction of Remijan et al., (2008), who used the data derived 
by Martin et al., (2004). 

\begin{figure}
\vskip 1cm
\centering
\includegraphics[height=8cm,width=8cm]{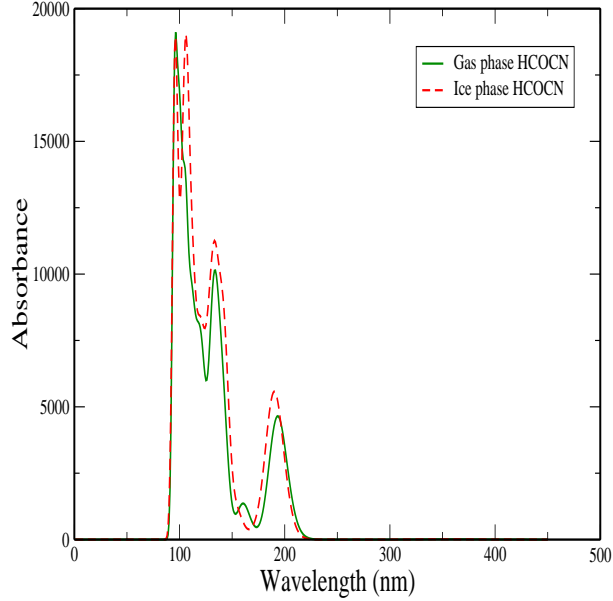}
\caption{Electronic absorption spectrum of HCOCN in gas well as in ice phase.}
\label{fig-2}
\end{figure}

To check the fractionation of HCOCN, we vary the initial gas phase atomic ratio of D/H (R$_D$)
of the cloud in the range from $0.001$ to $10$ by keeping all other initial parameters the 
same at n$_H=2 \times 10^4$ cm$^{-3}$, $T=16$K, A$_V$=10. In Fig. 5, we plot the 
DCOCN and HCOCN abundance ratio(peak and final values) in the 
ice phase with respect to the initial gas phase 
D/H ratio. As expected, the ratio increases with the increase of the 
atomic deuterium abundance in the gas phase. For the high value of initial gas phase D/H ratio, 
the fractionation of HCOCN ($>1$) exceeds the cosmic D/H ratio of $1.5 \times 10^{-5}$ 
(Roberts \& Millar 2000). From Fig.5, it is evident that the peak value of the fractionation ratio
greater than 1 for initial D/H ratio $>0.4$. Looking at the abundances of the DCOCN, we would
thus expect to observe DCOCN in the ISM as well.

\subsection{Spectral Analysis}

We now turn to the spectral properties. For this, we need to compute the infrared peak 
positions with their absorbance in the gas phase as well as in ice and mixed ice phase.
In Table 4, we present these for HCOCN and one of its isotopomers, namely, DCOCN. 
We find that the most intense mode of HCOCN in the gas phase 
appears nearly at 1655.43 cm$^{-1}$. This peak is shifted in the ice phase 
(water ice) by nearly 15 cm$^{-1}$ and appears at 1640.47 cm$^{-1}$. 
The second strongest peak in the gas phase 
appears at 938.68 cm$^{-1}$. It is also shifted in the ice phase. 
To have a more realistic condition, instead of only water ice, 
we consider a mixed ice mantle, which contains
70\% water, 20\% methanol and 10\% CO$_2$ molecules
(Keane et al., 2001; Das \& Chakrabarti, 2011). 
For the pure water ice, Gaussian 09W uses a dielectric constant of $\sim 78.5$ by default.
For the mixed ice, we put the dielectric constant
of the medium to be $61$, which is calculated by taking the weighted average of the dielectric
constants of H$_2$O, CH$_3$OH and CO$_2$. We note that
the most intense peak in the gas phase is shifted in the mixed ice also (Table 4). 
Isotope effects on the chemical shifts is caused by differences in 
vibrational modes due to the different isotope masses. 
Each of the HCOCN and DCOCN has a unique spectrum because the substitution of 
the isotope changes the reduced mass of the corresponding molecule.
We find that the most intense mode of DCOCN in the gas phase appears 
at 1635.75 cm$^{-1}$. This peak is shifted in the ice 
phase by 17.35 cm$^{-1}$, i.e., at 1618.37 cm$^{-1}$. 
The second strongest peak in the gas phase which appears at 888.25 cm$^{-1}$ 
is also shifted in the ice phase and appears at 879.59 cm$^{-1}$. 
The most intense peak in the gas phase is similarly shifted in the mixed solvated grain. 
Infrared peak positions with their absorbance in the gas phase as well as 
in the ice and mixed ice phase are pointed for the DCOCN in Table 4.
In Fig. 6, we show how the isotopic substitution (DCOCN) 
plays a part in the vibrational progressions of HCOCN in the ice phase. 
\begin{table*}
\scriptsize{
\centering
\vbox{
\caption{Vibrational frequencies of Cyanoformaldehyde molecule and one of its isotopomer 
in gas phase, H$_2$O ice, methanol ice and mixed water ice at B3LYP/6-311G++** 
level of theory}
\begin{tabular}{|c|c|c|c|c|c|c|}
\hline
{\bf Species}&{\bf Peak positions}&{\bf Absorbance}&{\bf Peak positions }&{\bf Absorbance}
&{\bf Peak positions}&{\bf Absorbance}\\
&{\bf (Gas phase)}&{}&{\bf (H$_2$O ice)}&{}&{\bf (Mixed ice) }&{}\\
&\bf (in cm$^{-1}$)&&\bf (in cm$^{-1}$)&&\bf ( in cm$^{-1}$)&\\
\hline
&237.18&12.69&238.11&21.92&238.32&21.00\\
&331.41&0.17&336.27&0.43&335.80&0.41\\
&648.66&1.88&649.40&1.90&649.59&1.90\\
&938.68&92.42&927.01&145.40&927.82&140.88\\
{\bf HCOCN}&1019.28&0.31&1022.82&0.84&1022.80&0.80\\
&1421.46&7.71&1421.99&16.15&1421.41&15.32\\
&1655.43&128.64&1640.47&223.04&	1641.74&214.43\\
&2269.21&31.24&2272.50&28.10&2272.22&29.39\\
&3033.11&33.56&3084.17&18.21&3080.19&19.37\\
\hline
&233.40&12.33&234.34&21.31&234.52&20.43\\
&315.40&1.53&319.79&2.78&319.39&2.66\\
&643.59&1.94&644.26&2.01&644.44&2.09\\
&876.02&0.59&876.81&0.53&877.00&0.54\\
{\bf DCOCN}&888.25&61.95&879.59&99.89&880.07&96.48\\
&1089.94&26.67&1087.86&42.18&1087.70&41.06\\
&1635.75&131.39&1618.37&230.78&1619.90&221.60\\
&2234.33&24.91&2260.81&18.19&2259.45&18.03\\
&2274.29&42.69&2290.78&39.28&2288.78&40.41\\
\hline
\end{tabular}}}
\end{table*}
In Table 5, we summarize the experimental values of the rotational and quartic
centrifugal distortion constants from Bogey et al. (1988) and our calculated
rotational and distortional constants of HCOCN molecule. Calculated constants are
corrected for each vibrational state as well as the vibrationally averaged structures.
Here we use MP2/aug-cc-pVTZ level to perform this calculations for the gas phase HCOCN molecule. 
In Table 5, calculated distortional constants along with the Experimental Quartic Centrifugal
distortion constants correspond to I$^t$ representation with `A' reduction are tabulated.
Calculated rotational constants (B and C) are in very good agreement with the measured values.
Our calculated values of B and C varies by 35 MHz and 29 MHz respectively from the experimentally
obtained B and C values. Other distortional constants($\Delta_J, \ \Delta_{JK}, \ \Delta_K, \ 
\delta_j, \ \delta_k$) are relatively closer to the experimentally obtained values.
Measured value of A deviates significantly from our calculated value.
This difference could be explained due to the difference between the techniques involved.
Calculation of rotational constants refer to an equilibrium geometry,
while the measured ones are subject to some vibrational effects also (Csaszar, 1989).
Calculated distortional constants in the asymmetrically reduced Hamiltonian
are very close to the measured quartic centrifugal rotational constants.
In order to summarize the outcome of our calculations about the rotational spectroscopy,
we have prepared our spectral information as per the guidelines of the JPL (Table 7 of Appendix A).

Different electronic absorption spectral parameters of HCOCN in the gas phase
are given in Table 6. In the gas phase, the spectrum is characterized by
four intense peaks at the wavelengths
$193.7$, $160.5$, $133.9$, $96.7$ nm (Fig. 7). These intense peaks are assigned
due to the Highest occupied molecular orbital (HOMO)- Lowest unoccupied molecular orbital (LUMO) 
transitions. These transitions correspond to
H-1$\rightarrow$ L+0, H-0$\rightarrow$ L+2, H-2$\rightarrow$ L+1 and
H-3$\rightarrow$ L+9.
Depending on the composition of the interstellar grain mantle
peak positions in the ice phase are shifted. These features are given
in the Table 6 with corresponding transition details. Fig. 7 clearly shows
some differences in the gas phase and ice phase electronic absorption spectra.


\begin{table*}
\scriptsize{
\centering
\vbox{
\caption{Theoretical \& Experimental rotational parameters of Cyanoformaldehyde molecule
}
\begin{tabular}{|c|c|c|c|c|c|c|}
\hline
{\bf Species}&{\bf Rotational }&{\bf Values}&
{\bf Experimental}&
{\bf Distortional }&{\bf Values}&{\bf Experimental} \\
&{\bf constants }&{\bf in MHz}&
{\bf  values}&
{\bf  constants }&{\bf in MHz}&{\bf values} \\
&&&{\bf in MHz$^a$}&&&{\bf in MHz$^a$} \\
\hline
&A&66034.4&67473.54&$\Delta_J$& 2.267$\times$10$^{-3}$&$2.266 \times 10^{-3}$\\
{\bf HCOCN in gas phase}&B&4975.9&5010.19&$\Delta_{JK}$&$-0.1413$&$-0.143104$\\
&C&4627.1&4656.498&$\Delta_{K}$&$7.074$&$8.99$\\
&&&&$\delta_j$&$3.933\times$10$^{-4}$&$3.877 \times 10^{-4}$\\
&&&&$\delta_k$&0.02895&$0.034325$\\
\hline
\multicolumn{2}{|c|}{$^a$ Bogey et al., (1988)}\\
\end{tabular}}}
\end{table*}


\begin{table*}
\scriptsize{
\centering
\caption{Electronic transitions of cyanoformaldehyde molecule at B3LYP/6-311++G** level}
\begin{tabular}{|c|c|c|c|c|c|}
\hline
{\bf Species}&{\bf Wavelength in nm}&{\bf Absorbance}&{\bf Oscillator strength }
&{\bf Transitions}&{\bf Contribution in \%}\\
\hline
&193.7&4651&0.1051&H-1$\rightarrow$ L+0&85\\
&160.5&1360&0.032&H-0$\rightarrow$ L+2&94\\
{\bf HCOCN}&133.9&10128&0.1234&H-2$\rightarrow$ L+1&63\\
&96.7&19111&0.0458&H-3$\rightarrow$ L+9&44\\
\hline
&190.7&5561&0.1307&H-1$\rightarrow$ L+0&85\\
&133.2&11260&0.1603&H-2$\rightarrow$ L+1&60\\
{\bf HCOCN in pure H$_2$O ice}&108.3&17562&0.0045&H-2$\rightarrow$ L+6&96\\
&95.3&18507&0.2275&H-0$\rightarrow$ L+13&34\\
\hline
&191.2&5618&0.1348&H-1$\rightarrow$ L+0&86\\
&135.4&11327&0.0528&H-0$\rightarrow$ L+5&92\\
{\bf HCOCN in mixed ice}&108.5&17667&0.0048&H-2$\rightarrow$ L+6&96\\
&98.1&16680&0.0289&H-4$\rightarrow$ L+5&89\\
\hline
\hline
\end{tabular}}
\end{table*}

\section{Conclusions}

Recently, there were some indications of observations of HCOCN molecules inside molecular clouds. In this
paper, we have investigated different aspects of the abundance of HCOCN in the ISM. They are:

\noindent {$\bullet$ We perform a quantum chemical calculation to find out the realistic rate 
coefficient for the formation of HCOCN molecule around the cold, dense cloud.}

\noindent{$\bullet$ Our chemical modeling shows that HCOCN and DCOCN could efficiently be formed 
in the ice phase. We are proposing that this molecule  may be observed in the ice phase as well.}

\noindent{$\bullet$ By introducing a large deuterated network into our chemical model, we have 
explored the possibility of studying one of the isotopologues (DCOCN) of HCOCN. We notice that
DCOCN is produced very efficiently in the ice phase. By varying the model parameters, 
we find that the fractionation ratio of HCOCN could be greater than 1 (Fig. 5).
Our Model calculation shows that maximum column density of ice phase HCOCN could be reached up to
$4.64 \times 10^{15}$ cm$^{-2}$ and for DCOCN it could be reached up to 
$3.34 \times 10^{14}$ cm$^{-2}$ in the density range of $2 \times 10^4 - 10^6$ cm$^{-3}$}.

\noindent {$\bullet$ We explore the vibrational and electronic 
spectral properties of HCOCN in different astrophysical conditions (such as, in the 
gas phase, ice phase and mixed ice phase). 
Different rotational \& distortional constants for the gas phase HCOCN are tabulated 
and compared with the experimentally obtained values. Moreover, in the Appendix A, we prepare a 
catalog file in JPL format for the gas phase HCOCN molecule which will be helpful for the observer.
Our result could be used as a guide to observers to look for this species in or around various molecular clouds.}

\noindent {$\bullet$ Based on certain observational results, we assumed grain mantles  of
different compositions. We considered grain mantles 
mainly composed of pure water. We also assumed the possibility that the ice 
mantle could be mixed in nature, consisting of water, methanol 
and carbon dioxide, for example. Based on the earlier studies, we assumed that
the interstellar grain mantle consists of 70\% water, 20\% Methanol 
and 10\% CO$_2$. We tabulated the differences in spectral properties 
due to the different types of ices. Future observations are expected to 
explore the composition of the ice mantle based on which our result may be revised further.}

\section{Acknowledgments}
We would like to thank the anonymous referees whose valuable suggestions have helped to
improve this paper a lot.
Ankan Das is grateful to ISRO for the financial support through a respond project 
(Grant No. ISRO/RES/2/372/11-12) and SC, SKC, RS, and LM thank a DST project 
(Grant No.SR/S2/HEP-40/2008).

\clearpage

\centering{\bf Appendix-A}
\clearpage
\begin{table*}
\caption{Different rotational transitions and its related parameters
for gas phase cyanoformaldehyde molecule in the format of JPL catalog.}
\begin{tabular}{|c|c|c|c|c|c|c|c|c|c|}
\hline
{\bf Frequency$^a$} & {\bf Uncertainty$^b$ } & {\bf I$^c$} & {\bf D$^d$ } & {\bf E$_{lower}$$^e$} & {\bf g$_{up}$$^f$ } & {\bf Tag$^g$} & {\bf QnF$^h$} & {\bf Qn$_{up}$$^i$} & {\bf Qn$_{lower}$$^j$}\\
\hline
    9602.0577 & 0.0000& -7.3358& 3  & -0.0000&  3 & 55001& 304& 1 0 1 1  &   0 0 0 1     \\
    9603.2441 & 0.0000& -7.1139& 3  & -0.0000&  5 & 55001& 304& 1 0 1 2  &   0 0 0 1    \\ 
    9605.0238 & 0.0000& -7.8128& 3  & -0.0000&  1 & 55001& 304& 1 0 1 0  &   0 0 0 1    \\ 
   19204.8519 & 0.0000& -7.0358& 3  &  0.3203&  5 & 55001& 304& 2 0 2 2  &   1 0 1 2    \\ 
   19205.0496 & 0.0000& -6.9108& 3  &  0.3204&  3 & 55001& 304& 2 0 2 1  &   1 0 1 0    \\ 
   19206.0382 & 0.0000& -6.5586& 3  &  0.3203&  5 & 55001& 304& 2 0 2 2  &   1 0 1 1    \\ 
   19206.1230 & 0.0000& -6.2875& 3  &  0.3203&  7 & 55001& 304& 2 0 2 3  &   1 0 1 2    \\ 
   19206.8293 & 0.0000& -8.2117& 3  &  0.3203&  3 & 55001& 304& 2 0 2 1  &   1 0 1 2    \\ 
   19208.0157 & 0.0000& -7.0357& 3  &  0.3203&  3 & 55001& 304& 2 0 2 1  &   1 0 1 1    \\ 
   28807.6502 & 0.0000& -6.8613& 3  &  0.9610&  7 & 55001& 304& 3 0 3 3  &   2 0 2 3    \\ 
   28808.7236 & 0.0000& -6.1289& 3  &  0.9610&  5 & 55001& 304& 3 0 3 2  &   2 0 2 1    \\ 
   28808.9213 & 0.0000& -5.9582& 3  &  0.9609&  7 & 55001& 304& 3 0 3 3  &   2 0 2 2    \\ 
   28808.9684 & 0.0000& -5.7979& 3  &  0.9610&  9 & 55001& 304& 3 0 3 4  &   2 0 2 3    \\ 
   28809.4299 & 0.0000& -8.4052& 3  &  0.9610&  5 & 55001& 304& 3 0 3 2  &   2 0 2 3    \\ 
   28810.7010 & 0.0000& -6.8613& 3  &  0.9609&  5 & 55001& 304& 3 0 3 2  &   2 0 2 2    \\ 
   38410.3229 & 0.0000& -6.7387& 3  &  1.9219&  9 & 55001& 304& 4 0 4 4  &   3 0 3 4    \\ 
   38411.5564 & 0.0000& -5.6807& 3  &  1.9220&  7 & 55001& 304& 4 0 4 3  &   3 0 3 2    \\ 
   38411.6412 & 0.0000& -5.5626& 3  &  1.9219&  9 & 55001& 304& 4 0 4 4  &   3 0 3 3    \\ 
   38411.6711 & 0.0000& -5.4474& 3  &  1.9219& 11 & 55001& 304& 4 0 4 5  &   3 0 3 4    \\ 
   38412.0179 & 0.0000& -8.5379& 3  &  1.9219&  7 & 55001& 304& 4 0 4 3  &   3 0 3 4    \\ 
   38413.3361 & 0.0000& -6.7387& 3  &  1.9219&  7 & 55001& 304& 4 0 4 3  &   3 0 3 3    \\ 
   48012.7951 & 0.0000& -6.6448& 3  &  3.2032& 11 & 55001& 304& 5 0 5 5  &   4 0 4 5    \\ 
   48014.0962 & 0.0000& -5.3560& 3  &  3.2032&  9 & 55001& 304& 5 0 5 4  &   4 0 4 3    \\ 
   48014.1433 & 0.0000& -5.2646& 3  &  3.2032& 11 & 55001& 304& 5 0 5 5  &   4 0 4 4    \\ 
   48014.1641 & 0.0000& -5.1743& 3  &  3.2032& 13 & 55001& 304& 5 0 5 6  &   4 0 4 5    \\ 
   48014.4430 & 0.0000& -8.6403& 3  &  3.2032&  9 & 55001& 304& 5 0 5 4  &   4 0 4 5    \\ 
   48015.7912 & 0.0000& -6.6448& 3  &  3.2032&  9 & 55001& 304& 5 0 5 4  &   4 0 4 4    \\ 
   57615.0045 & 0.0000& -6.5693& 3  &  4.8048& 13 & 55001& 304& 6 0 6 6  &   5 0 5 6    \\ 
   57616.3435 & 0.0000& -5.1002& 3  &  4.8048& 11 & 55001& 304& 6 0 6 5  &   5 0 5 4    \\ 
   57616.3734 & 0.0000& -5.0252& 3  &  4.8048& 13 & 55001& 304& 6 0 6 6  &   5 0 5 5    \\ 
   57616.3886 & 0.0000& -4.9509& 3  &  4.8048& 15 & 55001& 304& 6 0 6 7  &   5 0 5 6    \\ 
   57616.6224 & 0.0000& -8.7245& 3  &  4.8048& 11 & 55001& 304& 6 0 6 5  &   5 0 5 6    \\ 
   57617.9913 & 0.0000& -6.5693& 3  &  4.8048& 11 & 55001& 304& 6 0 6 5  &   5 0 5 5    \\ 
   67216.8929 & 0.0000& -6.5067& 3  &  6.7267& 15 & 55001& 304& 7 0 7 7  &   6 0 6 7    \\ 
   67218.2563 & 0.0000& -4.8891& 3  &  6.7267& 13 & 55001& 304& 7 0 7 6  &   6 0 6 5    \\ 
   67218.2770 & 0.0000& -4.8255& 3  &  6.7266& 15 & 55001& 304& 7 0 7 7  &   6 0 6 6    \\ 
   67218.2887 & 0.0000& -4.7622& 3  &  6.7267& 17 & 55001& 304& 7 0 7 8  &   6 0 6 7    \\ 
   67218.4900 & 0.0000& -8.7966& 3  &  6.7267& 13 & 55001& 304& 7 0 7 6  &   6 0 6 7    \\ 
   67219.8742 & 0.0000& -6.5067& 3  &  6.7266& 13 & 55001& 304& 7 0 7 6  &   6 0 6 6    \\ 
   76818.4039 & 0.0000& -6.4537& 3  &  8.9688& 17 & 55001& 304& 8 0 8 8  &   7 0 7 8    \\ 
   76819.7845 & 0.0000& -4.7097& 3  &  8.9688& 15 & 55001& 304& 8 0 8 7  &   7 0 7 6    \\ 
   76819.7997 & 0.0000& -4.6544& 3  &  8.9688& 17 & 55001& 304& 8 0 8 8  &   7 0 7 7    \\ 
   76819.8089 & 0.0000& -4.5993& 3  &  8.9688& 19 & 55001& 304& 8 0 8 9  &   7 0 7 8    \\ 
   76819.9859 & 0.0000& -8.8601& 3  &  8.9688& 15 & 55001& 304& 8 0 8 7  &   7 0 7 8    \\ 
   76821.3816 & 0.0000& -6.4537& 3  &  8.9688& 15 & 55001& 304& 8 0 8 7  &   7 0 7 7    \\ 
   86419.4821 & 0.0000& -6.4083& 3  & 11.5313& 19 & 55001& 304& 9 0 9 9  &   8 0 8 9    \\ 
   86420.8754 & 0.0000& -4.5541& 3  & 11.5313& 17 & 55001& 304& 9 0 9 8  &   8 0 8 7    \\ 
   86420.8871 & 0.0000& -4.5052& 3  & 11.5312& 19 & 55001& 304& 9 0 9 9  &   8 0 8 8    \\ 
   86420.8945 & 0.0000& -4.4563& 3  & 11.5313& 21 & 55001& 304& 9 0 910  &   8 0 8 9    \\ 
   86421.0524 & 0.0000& -8.9173& 3  & 11.5313& 17 & 55001& 304& 9 0 9 8  &   8 0 8 9    \\ 
   86422.4574 & 0.0000& -6.4083& 3  & 11.5312& 17 & 55001& 304& 9 0 9 8  &   8 0 8 8    \\ 
   96020.0723 & 0.0000& -6.3689& 3  & 14.4140& 21 & 55001& 304&10 01010  &   9 0 910    \\ 
   96021.4755 & 0.0000& -4.4172& 3  & 14.4140& 19 & 55001& 304&10 010 9  &   9 0 9 8    \\ 
   96021.4847 & 0.0000& -4.3732& 3  & 14.4139& 21 & 55001& 304&10 01010  &   9 0 9 9    \\ 
   96021.4908 & 0.0000& -4.3294& 3  & 14.4140& 23 & 55001& 304&10 01011  &   9 0 910    \\ 
   96021.6334 & 0.0000& -8.9697& 3  & 14.4140& 19 & 55001& 304&10 010 9  &   9 0 910    \\ 
\hline
\end{tabular}
\end{table*}

\begin{table*}
\begin{tabular}{|c|c|c|c|c|c|c|c|c|c|}
\hline
{\bf Frequency$^a$} & {\bf Uncertainty$^b$ } & {\bf I$^c$} & {\bf D$^d$ } & {\bf E$_{lower}$$^e$} & {\bf g$_{up}$$^f$ } & {\bf Tag$^g$} & {\bf QnF$^h$} & {\bf Qn$_{up}$$^i$} & {\bf Qn$_{lower}$$^j$}\\
\hline
   96023.0458 & 0.0000& -6.3689& 3  & 14.4139& 19 & 55001& 304&10 010 9  &   9 0 9 9    \\ 
  105620.1196 & 0.0000& -6.3345& 3  & 17.6169& 23 & 55001& 304&11 01111  &  10 01011    \\ 
  105621.5307 & 0.0000& -4.2952& 3  & 17.6169& 21 & 55001& 304&11 01110  &  10 010 9    \\ 
  105621.5382 & 0.0000& -4.2553& 3  & 17.6168& 23 & 55001& 304&11 01111  &  10 01010    \\ 
  105621.5433 & 0.0000& -4.2155& 3  & 17.6169& 25 & 55001& 304&11 01112  &  10 01011    \\ 
  105621.6733 & 0.0000& -9.0183& 3  & 17.6169& 21 & 55001& 304&11 01110  &  10 01011    \\ 
  105623.0918 & 0.0000& -6.3345& 3  & 17.6168& 21 & 55001& 304&11 01110  &  10 01010    \\ 
  115219.5693 & 0.0000& -6.3044& 3  & 21.1400& 25 & 55001& 304&12 01212  &  11 01112    \\ 
  115220.9869 & 0.0000& -4.1855& 3  & 21.1400& 23 & 55001& 304&12 01211  &  11 01110    \\ 
  115220.9930 & 0.0000& -4.1491& 3  & 21.1400& 25 & 55001& 304&12 01212  &  11 01111    \\ 
  115220.9974 & 0.0000& -4.1126& 3  & 21.1400& 27 & 55001& 304&12 01213  &  11 01112    \\ 
  115221.1169 & 0.0000& -9.0639& 3  & 21.1400& 23 & 55001& 304&12 01211  &  11 01112    \\ 
  115222.5406 & 0.0000& -6.3044& 3  & 21.1400& 23 & 55001& 304&12 01211  &  11 01111    \\ 
  124818.3668 & 0.0000& -6.2780& 3  & 24.9834& 27 & 55001& 304&13 01313  &  12 01213    \\ 
  124819.7898 & 0.0000& -4.0863& 3  & 24.9834& 25 & 55001& 304&13 01312  &  12 01211    \\ 
  124819.7949 & 0.0000& -4.0527& 3  & 24.9834& 27 & 55001& 304&13 01313  &  12 01212    \\ 
  124819.7987 & 0.0000& -4.0191& 3  & 24.9834& 29 & 55001& 304&13 01314  &  12 01213    \\ 
  124819.9092 & 0.0000& -9.1072& 3  & 24.9834& 25 & 55001& 304&13 01312  &  12 01213    \\ 
  124821.3373 & 0.0000& -6.2780& 3  & 24.9834& 25 & 55001& 304&13 01312  &  12 01212    \\ 
  134416.4575 & 0.0000& -6.2548& 3  & 29.1469& 29 & 55001& 304&14 01414  &  13 01314    \\ 
  134417.8850 & 0.0000& -3.9960& 3  & 29.1469& 27 & 55001& 304&14 01413  &  13 01312    \\ 
  134417.8894 & 0.0000& -3.9648& 3  & 29.1469& 29 & 55001& 304&14 01414  &  13 01313    \\ 
  134417.8927 & 0.0000& -3.9336& 3  & 29.1469& 31 & 55001& 304&14 01415  &  13 01314    \\ 
  134417.9955 & 0.0000& -9.1485& 3  & 29.1469& 27 & 55001& 304&14 01413  &  13 01314    \\ 
  134419.4274 & 0.0000& -6.2548& 3  & 29.1469& 27 & 55001& 304&14 01413  &  13 01313    \\ 
  144013.7869 & 0.0000& -6.2346& 3  & 33.6306& 31 & 55001& 304&15 01515  &  14 01415    \\ 
  144015.2183 & 0.0000& -3.9134& 3  & 33.6306& 29 & 55001& 304&15 01514  &  14 01413    \\ 
  144015.2221 & 0.0000& -3.8843& 3  & 33.6306& 31 & 55001& 304&15 01515  &  14 01414    \\ 
  144015.3211 & 0.0000& -9.1882& 3  & 33.6306& 29 & 55001& 304&15 01514  &  14 01415    \\ 
  144016.7563 & 0.0000& -6.2346& 3  & 33.6306& 29 & 55001& 304&15 01514  &  14 01414    \\ 
  153611.7352 & 0.0000& -3.8376& 3  & 38.4345& 31 & 55001& 304&16 01615  &  15 01514    \\ 
  153613.2694 & 0.0000& -6.2169& 3  & 38.4344& 31 & 55001& 304&16 01615  &  15 01515    \\ 
\end{tabular}
\end{table*}
\clearpage
\footnote{
$^a$ Calculated frequency in MHz\\
$^b$ Calculated uncertainty of the line. If the line position is in units of MHz then uncertainty of the 
line is greater or equal to zero.\\
$^c$ Base 10 logarithm of the integrated intensity at 300K in nm$^2$ MHz\\
$^d$ Degrees of freedom in the rotational partition function (0 for atoms, 2 for linear molecules, 3 for non linear molecules)\\
$^e$ Lower state energy in cm$^{-1}$ relative to the lowest energy level in the ground vibrionic state. \\
$^f$ Upper state degeneracy : g$_{up}=g_{I} \times g_{N}$, where g$_{I}$ is the spin statistical weight and g$_{N} =2N+1$ the rotational degeneracy.\\
$^g$ Molecule Tag\\
$^h$ Coding for the format of quantum numbers. QnF=$100 \times Q + 10 \times H + N_{Qn}$; N$_{Qn}$ is the number of quantum numbers for each state;
H indicates the number of half integer quantum numbers; Qmod5, the residual when Q is divided by 5, gives the number of principal
quantum numbers (without the spin designating ones).\\
$^i$ Quantum numbers for the upper state\\
$^j$ Quantum numbers for the lower state}
\end{document}